# Unambiguous Determination of the Spin of the Black Hole in NGC 1365

G. Risaliti[1,2], F. A. Harrison[3], K. K. Madsen[3], D. J. Walton[3], S. E. Boggs[4], F. E. Christensen[5], W. W. Craig[5,6], B. W. Grefenstette[3], C. J. Hailey[7], E. Nardini[8], Daniel Stern[9], W. W. Zhang[10]

[1]INAF- Osservatorio Astrofisico di Arcetri, L.go E. Fermi 5 50125 Firenze, IT
[2]Harvard-Smithsonian Center for Astrophysics, 60 Garden Street, Cambridge, MA 02138, USA
[3]Cahill Center for Astrophysics, 1216 E. California Blvd, California Institute of Technology, Pasadena, CA 91125, USA
[4]Space Science Laboratory, University of California, Berkeley, CA 94720, USA
[5]Danish Technical University, Lyngby, DK
[6]Lawrence Livermore National Laboratory, Livermore, CA, USA
[7]Columbia University, New York, NY 10027, USA
[8]Astrophysics Group, School of Physical and Geographical Sciences, Keele University, Keele, Staffordshire ST5 5BG, UK
[9]Jet Propulsion Laboratory, California Institute of Technology, 4800 Oak Grove Drive, Pasadena, CA 91109
[10]NASA Goddard Space Flight Center, Greenbelt, MD 20771, USA

**Broad X-ray emission lines from neutral and partially ionized iron observed in active galaxies have been interpreted as fluorescence produced by the reflection of hard X-rays off the inner edge of an accretion disk[1,2,3,4,5,6,7]. In this model, line broadening and distortion result from rapid rotation and relativistic effects near the black hole, the line shape being sensitive to its spin. Alternative models where the distortions result from absorption by intervening structures provide an equally good description of the data[8,9], and there has been no general agreement on which is correct. Recent claims[10] that**

**the M~2 x 10$^6$ M$_{SUN}$ black hole[11,12] at the center of NGC 1365 is close to maximally rotating rest upon the assumption of relativistic reflection. Here we report broadband X-ray observations of NGC 1365 that reveal the relativistic disk features through broadened Fe-line emission and an associated 10- 30 keV Compton scattering excess. Using temporal and spectral analyses we disentangle continuum changes due to time variable absorption from reflection, which we find arises from a region within 2.5 gravitational radii of a rapidly spinning black hole. Absorption dominated models that do not include relativistic disk reflection can be ruled out both statistically and on physical grounds.**

Previous soft (E<10 keV) X-ray observations of NGC 1365 have consistently seen Compton-thin clouds that periodically cross the line of sight, absorbing the 3 – 5 keV flux, and changing the continuum on timescales of hours[13,14]. On timescales of a day the spectrum has been seen to change from a state where a non-thermal continuum (associated with a tenuous coronal plasma) is viewed through a moderate, variable obscuring column to a state where the emission is completely dominated by Compton reflection off a dense material[15]. Based on the black hole mass, and assuming velocities of a few 10$^3$ km/s for the obscuring clouds, the inferred size of the X-ray source is a few gravitational radii[13].

XMM-Newton[16] and NuSTAR[17] simultaneously observed NGC 1365 with an exposure of ~130 ksec from July 25 – 27 (UT), 2012. These broadband X-ray data (0.5 – 79 keV) have sufficient statistics to study spectral variability on the relevant few-hour timescales. We find the low energy (E < 3 keV) component to be constant, and dominated by thermal emission from an optically-thin plasma (see SI). Previous Chandra imaging observations show this soft component is spatially extended, and physically distinct from the variable emission[18]. We restrict our analysis to the hard components observed above 3 keV.

The 3-79 keV spectrum shows strong emission features typical of relativistic disk reflection (Figure 1). However, previous work has demonstrated that the 5-7 keV distortion can be also be explained by varying absorbers along the line of sight partially covering the source. These have been seen before in NGC 1365, and they strongly affect the spectral shape below 10 keV[10,13].

In the 3–10 keV XMM-Newton band, the flux ratio F(3-5 keV)/F(6-10 keV) shows strong time variations, while the F(7-15 keV)/F(15-80 keV) ratio in the NuSTAR band is constant over the whole observation (Figure 2), suggesting that the soft variation is due to variable absorption (see SI). This enables us to perform a time-resolved analysis and accurately decompose the different spectral components. We break the observation into four intervals of similar hardness and analyze each simultaneously. We consider two models to explain the spectrum and its variability, one containing a relativistic reflection component from the inner accretion disk and a variable, partially covering absorber (model 1), and one employing multiple variable absorbers instead (model 2). In addition both models include neutral reflection from a constant, distant screen of gas to reproduce the unresolved, narrow iron emission line at 6.4 keV.

To illustrate the power of the broad energy range in determining model parameters, we first fit the XMM data alone to both models. With XMM data alone, we cannot distinguish between these scenarios, the broadened Fe-line being consistent with both. However, the extrapolation of the two models to the 10-80 keV range breaks the degeneracy: the absorption-only model fails to reproduce the new NuSTAR data by a large factor, while the model including the relativistic component reproduces the observed spectrum within ~5%, without even re-fitting the data (Figure 3).

We next consider the possibility that the shape of the hard spectrum arises not from disk reflection, but from an interleaving screen of dense absorbing material that can scatter a fraction of the light, but primarily affects the high-energy continuum through absorption. High density (far in excess of that found by the fits to XMM-only) is required to affect energies above 10 keV. We fit the broad 3- 79 keV band allowing the absorber parameters in model 2 to vary relative to those found with XMM alone. The resulting fit drives the column of the second absorber to $5 \times 10^{24}$ cm$^{-2}$, which reproduces the shape of the hard continuum, but cannot reproduce the spectral variability below 10 keV (see SI). We next add a third absorber to model 2, and allow the parameters of all three to vary. If the third absorber has $N_H = 4 – 6 \times 10^{24}$ cm$^{-2}$ and covers at least 40% of the source, we can achieve a formally acceptable fit (reduced $\chi^2$ of 1.02). However, this solution is rejected compared with the reflection model: an F-test performed on a ``merged model'' that includes all the components of the two individual models shows that the probability that the model without relativistic reflection is preferred is $8 \times 10^{-5}$.

The three-zone model can also be ruled out based on physical implications that are completely inconsistent with optical, IR and X-ray observations. These arise when we consider the strong Compton scattering and reprocessing inevitable with the high column density absorber. For low-column ($N_H < 10^{24}$ cm$^{-2}$) absorbers the effect of continuum reprocessing can be ignored (and in fact is not included in our models). However, the high-density absorber required to fit the E>10 keV data would strongly affect the predicted line and continuum emission. How strongly depends on the geometry. We consider two extreme bounding possibilities. First, if the solid angle covered is small, the reprocessing into X-ray line emission or infrared flux will be negligible. But, in this case the effect of Compton scattering on the direct emission will strongly (by a factor $\exp(\tau_C) \sim 20-80$) attenuate the intrinsic luminosity. Correcting for this, the implied total 0.1–100 keV luminosity becomes implausibly high ($L_x \sim 1\text{-}3 \times 10^{44}$ erg s$^{-1}$, i.e. 30-100% of Eddington). This is completely incompatible with other luminosity indicators, such as the optical [O III] 5007 A line[20]. Alternately, if the Compton-thick absorber covers a large fraction of the solid angle as seen from the source, there will be strong reprocessing, with two observable effects: a large fraction of the intrinsic luminosity is re-emitted in the infrared, which is ruled out by high-resolution IR observations[21], and a strong narrow iron emission line[22] with equivalent width EW>600 eV should be present, which is ruled out by our measurement - $EW_{OBS}\sim 60$ eV (see SI).

The observed source variability and unprecedented broad-band spectroscopy enable us to unambiguously conclude that reflection of a hard continuum off the inner accretion disk edge is producing the neutral Fe-line distortion and strong high-energy Compton reflection continuum.  The flux of the relativistic reflection component is higher than the primary continuum at its peak around 30 keV (see SI), implying a strong enhancement of the reflection efficiency due to relativistic distortion[23] (the highest value with standard disk reflection is ~30%[24]). This high efficiency is consistent with the disk and spin parameters presented below, implying that most of the reflection arises from the inner part of the accretion disk, close to the Innermost Stable Circular Orbit  (ISCO).

From the relativistic disc reflection, we can determine the black hole spin parameter. Allowing all other model parameters (e.g. disk inclination, ionization state, emissivity profile) to vary, we find the minimum spin to be $a^* \geq 0.84$ at 90% confidence (Fig. 4), where $a^*$ is the dimensionless spin parameter, $a^*=Jc/GM^2$, equivalent to an innermost disk edge at ≤2.5 gravitational radii. These results are consistent with the ones from previous Suzaku observations of NGC 1365 (REF 10). In that analysis, however, the relativistic model was assumed to be valid, and a complete comparison with the absorption-only scenario was not attempted, due to the low statistics, especially at high energies (see SI for further details).  Having tested this model against the possible alternative (i.e. the multiple variable absorber), and having left all the model parameters unconstrained, the main potential source of residual systematic error is due to assuming truncation of the disk at the ISCO.  However, recent MHD simulations suggest that emission from within the ISCO is negligible[26].  Other errors could be introduced by assuming a constant, power law, emissivity profile, however the steepness of this profile suggests that its exact shape at radii larger than twice the ISCO is unimportant.   The analysis therefore provides robust support for high spin values in AGN[2], constraining galaxy evolution and black hole growth models[27].

**Acknowledgements** This work was supported under NASA No. NNG08FD60C, and made use of data from the Nuclear Spectroscopic Telescope Array (NuSTAR) mission, a project led by Caltech, managed by the Jet Propulsion Laboratory, and funded by the National Aeronautics and Space Administration. We thank the NuSTAR Operations, Software and Calibration teams for support with execution and analysis of these observations. This work also made use of observations obtained with XMM-Newton, an ESA science mission with instruments and contributions directly funded by ESA Member States and NASA.

**Author Contributions** G.R. is PI of the XMM observations of NGC1365. He led the XMM data analysis and joint spectral analysis, modeling and interpretation of the results. F.A.H. is Principal Investigator of NuSTAR and led observation planning, execution, and participated in scientific interpretation and manuscript preparation. D.W. analyzed NuSTAR data and participated in modeling and interpretation. K.K.M. led NuSTAR calibration analysis for NGC1365, D.S. is NuSTAR Project Scientist and participated in definition and interpretation of the observations. B.G. assisted in adaptation of the NuSTAR data analysis to NGC1365. E.N. participated in XMM observations, reduction and interpretation. S.B., F.E.C., W.C., C.J.H and W.W.Z led efforts in design, calibration, and implementation of NuSTAR. All authors participated in review of the manuscript.

**Author Information** Reprints and permissions information is available at www.nature.com/reprints. Correspondence and requests for materials should be addressed to risaliti@arcetri.astro.it and fiona@srl.caltech.edu

**Figure 1** The broadband 3 – 79 keV X-ray spectrum of NGC 1365. An absorbed power law has been fit to the usually featureless continuum intervals 3-4 keV, 7-10 keV, and 50-80 keV. The residuals relative to this fit are shown for NuSTAR modules A and B (black and red data in main panel), and for the XMM PN (inset panel). Three prominent features are apparent: an asymmetric excess between 5 and 7 keV, a broad prominent excess between 10 and 79 keV, and a series of absorption lines between 6.7-8.3 keV. The latter are due to absorption by a highly ionized plasma[19], which we include in all the spectral models discussed below. The two former are typical signatures of relativistically blurred emission. Quoted errors refer to a 90% confidence level.

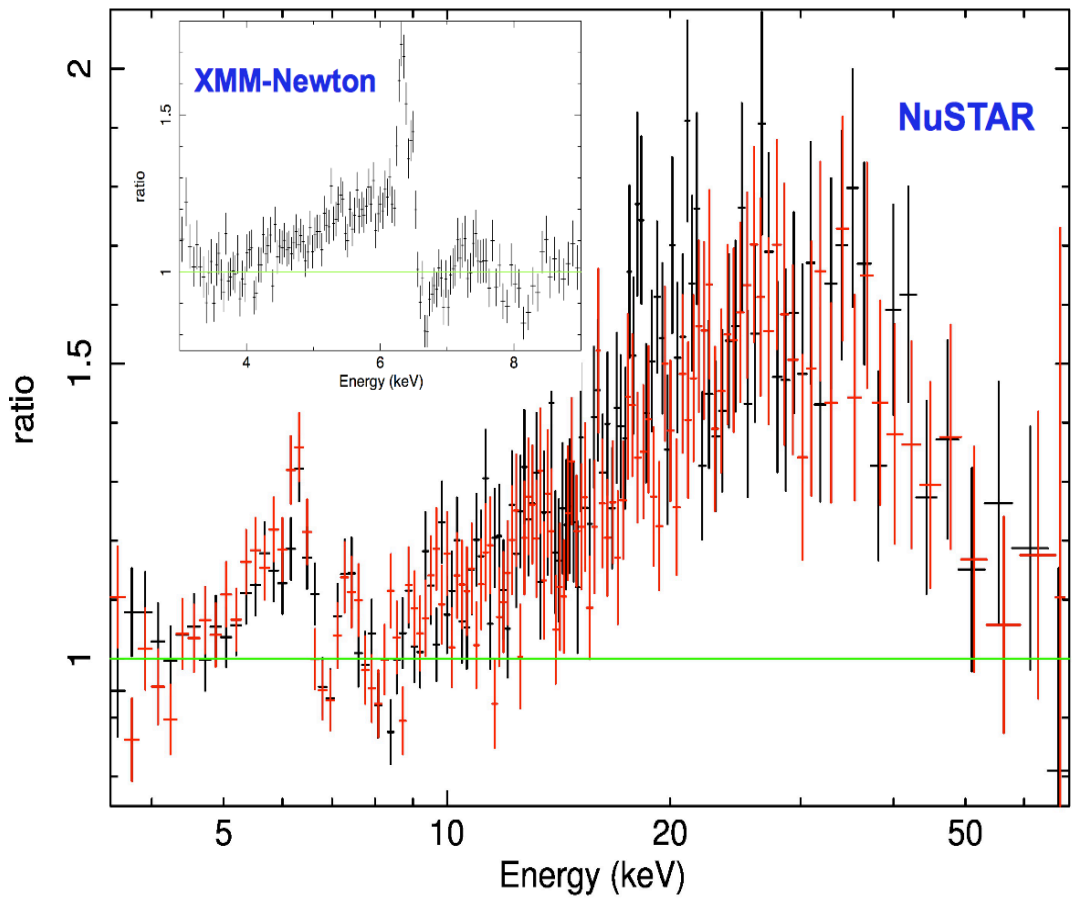

**Figure 2** X-ray spectral variability of NGC 1365. Upper panel: NuSTAR F(7-15 keV)/F(15-80 keV) Softness Ratio (SR) light curve. Lower panel: XMM F(3-5 keV)/F(6-10 keV) Softness Ratio (SR) light curve. The vertical lines delimit four time intervals with significanlty different average values of SR, from which four different spectra have been extracted and analyzed simultaneously. These variations are due to absorption variability (see SI). The changes of SR within each interval reveal a complex structure of the absorber; however its detailed variability on time scales shorter than ~$10^3$ s is not relevant in the determination of the physical parameters of the other spectral components. Error bars are one standard deviation.

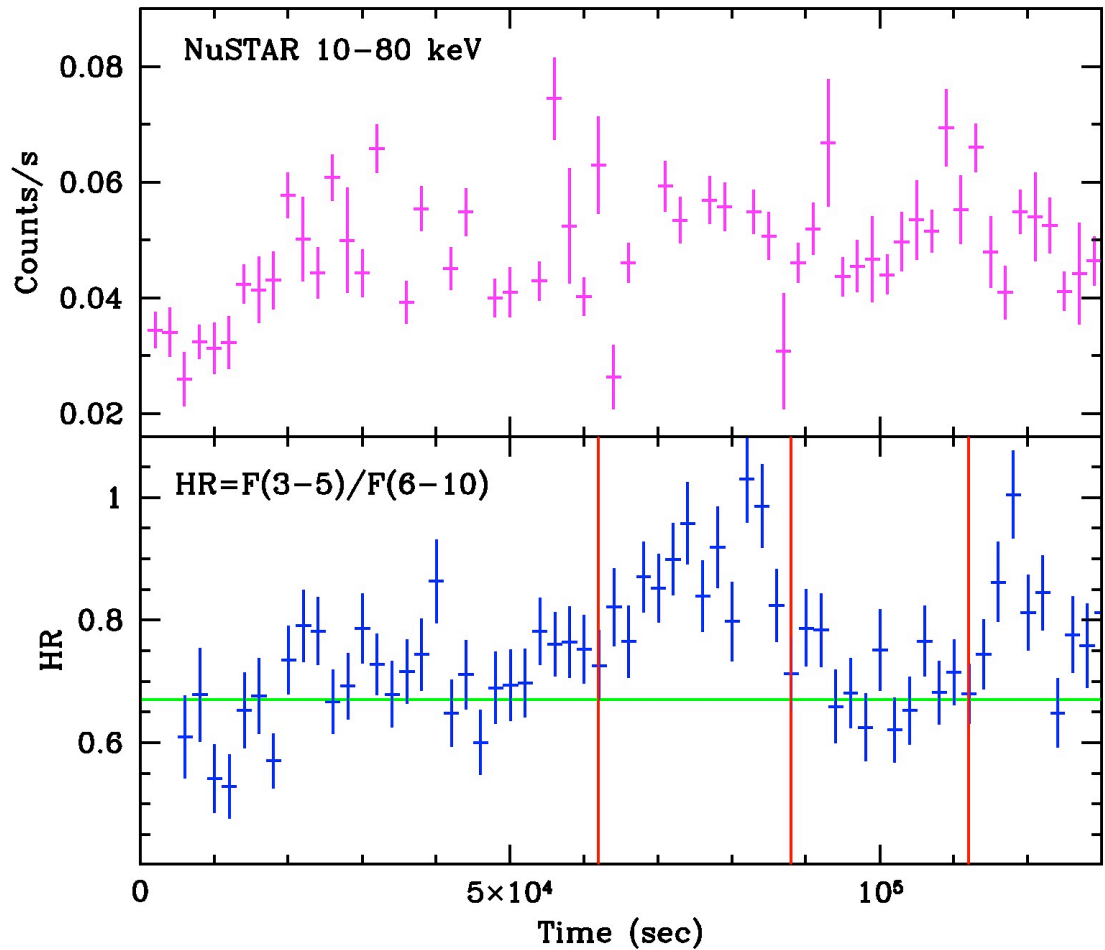

**Figure 3** Comparison between the relativistic reflection model and the multiple absorber model. Upper panel: XMM-Newton and NuSTAR spectral data+models for one of the four time intervals in Fig. 2. The two models contain (a) a relativistic reflection component plus variable partial covering (red dashed line), and (b) a double partial covering (blue continuous line). Both models have been fitted to the data only below 10 keV, and reproduce the lower energy observations equally well. However, the two models strongly deviate at higher energies. Lower panel: data/model ratio for the double partial covering (blue) and relativistic reflection+variable absorber (red) models. The best-fit partial covering model using XMM data alone has two absorbers with columns of $5\times10^{22}$ cm$^{-2}$ and $3\times10^{23}$ cm$^{-2}$; these have little influence on the spectrum above 10 keV. Instead, the relativistic reflection component reproducing the broad emission feature below 10 keV also includes a strong Compton reflection component above 10 keV. All errors are at a 90% confidence level.

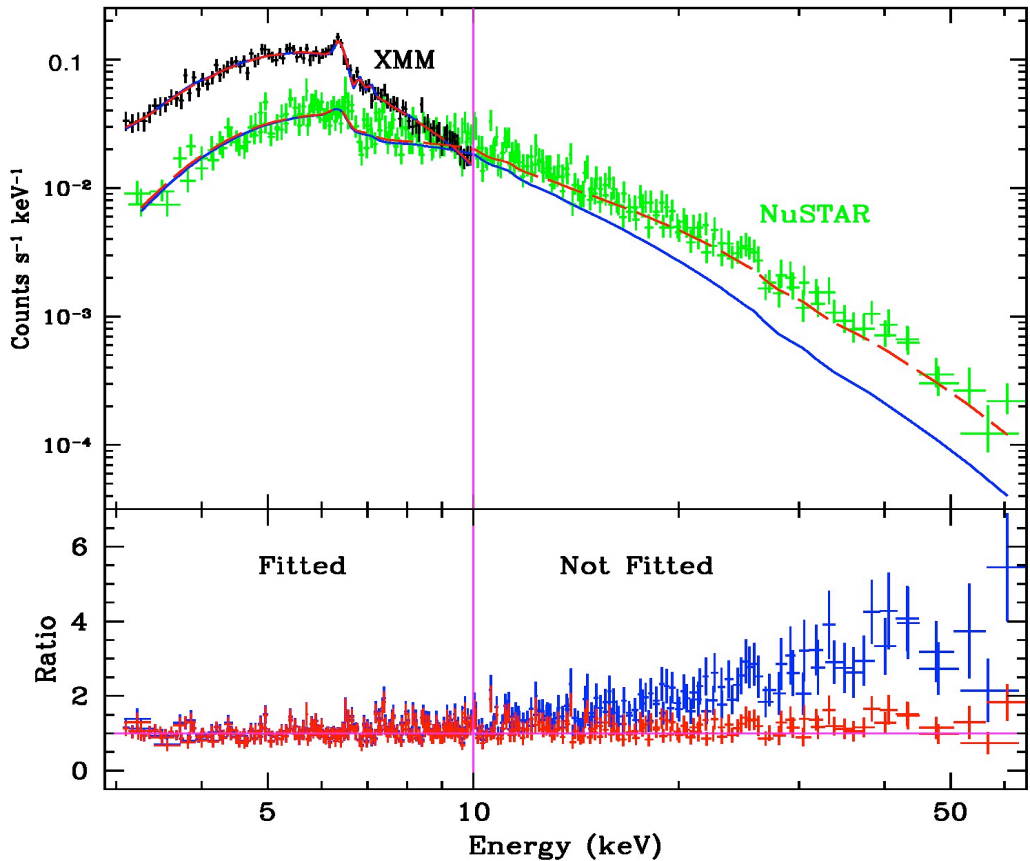

**Figure 4** Error contour for the spin parameter of the supermassive black hole in NGC 1365. The $\chi^2$ contour has been obtained allowing the variability of all model parameters. The adopted best fit model consists of a power law with $\Gamma=1.92^{+0.04}_{-0.26}$, a neutral absorber with column density varying in the range 2.2-2.8×10$^{23}$ cm$^{-2}$, a ionized absorber with ionization parameter ~3.5 and column density ~10$^{23}$ cm$^{-2}$, a relativistically blurred reflection by a disk with inner radius R<2.5 R$_G$ and spin parameter as shown in the plot. The ionization state of the disk, its inclination, its iron abundance and the normalization of the reflection component are degenerate, and so are only loosely constrained; they also strongly affect the uncertainties on the spin parameter: to emphasize the importance of correctly considering systematic effects, if based on previous work on Suzaku data[25] we limit the disk inclination to 55-60 degrees, the error on the spin measurement drops to a*=0.97$^{+0.01}_{-0.04}$. Analogously, we would obtain an error of the order of 0.01 if we fit the whole data set with a model not allowing for absorption variability.

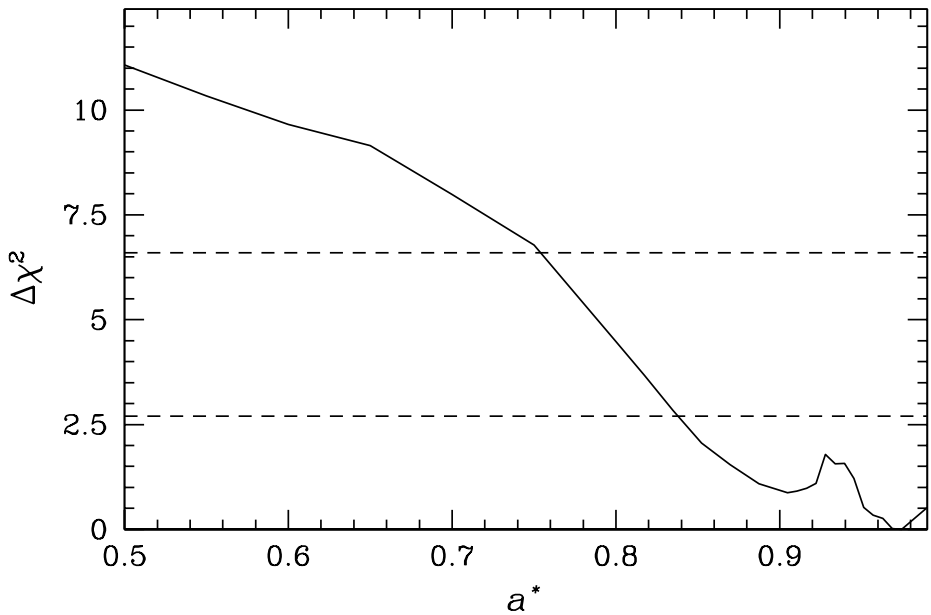

**Supplementary Information**

**Data reduction:**
NGC 1365 was observed simultaneously by XMM-Newton and NuSTAR on July 25-27, 2012 for a total integration time of ~130 ks.

We reduced and calibrated the XMM-Newton data following the standard procedures making use of the SAS 9.0 software provided by the XMM-Newton Data Center at ESAC, Spain. The EPIC data were checked for high-background flares, which may affect the spectral quality by reducing the signal-to-noise ratio (S/N). We did not detect any strong flares for the first ~120 ks, while the final 15 ks are affected by higher-than-average background activity. To be conservative we neglected this time interval in our subsequent study. We extracted the background from two different regions close to the target and compared them in order to check for systematic differences. We found the results to be consistent. Moreover, the total background flux is less than 2%, and not higher than 5%, of the total source flux at any energy in the interval 0.5-10 keV. We repeated the same procedure for the EPIC PN, MOS1 and MOS2. The results for the two MOS units are identical within statistical measurement errors, and so we merged the final products and calibrations. The spectral analysis was performed simultaneously on the PN and merged MOS data sets, allowing for a free cross-calibration constant between the two instruments.

The NuSTAR data have been reduced using the NuSTAR Data Analysis Software (NUSTARDAS) version 6.1. The NUSTARDAS pipeline software is fully HEASARC ftool compatible and is written and maintained by the ASI Science Data Center. For the NuSTAR response we used the NuSTAR caldb V20120918, maintained and updated by the NuSTAR Science Operations Center at Caltech. The data have been cleaned for SAA passages and reduced with the default depth correction, which significantly decreases the internal background at high energies. The background was extracted from an equivalently-sized region on the same detector, as close as possible to the source without being contaminated by the PSF wings. We estimate that the error in the background for this source extraction is less than 0.5%. The extraction regions and the source and background spectra are shown in Figure SI 1 and SI 2, together with a comparison with imaging and spectral capabilities of other currently available instruments at energies E>10 keV.

At the time of writing the NuSTAR vignetting function was not incorporated into the standard response files, which have a ~5% difference between the two telescope modules for some ranges of off-axis angle, as well as corrections to the response for energies below 5 keV that have not been incorporated into the standard pipeline. In order to quantitatively assess the contribution of residual calibration uncertainties present in this version of the calibration database , we used a Crab observation obtained at the same off-axis angle to evaluate the responses relative to the pre-launch optics model. For the off-axis angles sampled in this observation, the Crab calibration data demonstrate any residual response errors are at <2%. These corrections are much smaller than the variations in the physical models presented in the paper, and we have checked that they do not affect the parameter measurements in any of our fits. In order to minimize systematic effects, we do not combine responses or spectra from the two modules, but fit them simultaneously with no loss in statistical precision. We allow for free cross-calibrations between each of the NuSTAR modules and XMM-Newton. The absolute cross calibration factor is small (<3%) and we obtain excellent agreement in the expected spectral shape below 10 keV, where there is good overlap in responses. We do not include NuSTAR data below 5 keV in the model fits in order to avoid possible residual low-energy response uncertainties. Given XMM's larger collecting area below 5 keV this does not degrade the statistical model constraints.

**Data Analysis:**
**1. Soft emission.** The total 0.5-10 keV spectrum obtained from the XMM-Newton observation is shown in Fig. SI 3. The 0.5-2.5 keV emission has been studied in detail in past work[18,28]. It arises from a two-component optically thin plasma, extended on scales of ~2 kpc with temperatures $kT_1$~0.3 keV and $kT_2$~0.8 keV. The steep decrease of this emission at energies E>2 keV, and the photoelectric cut-off in the higher energy spectrum make the soft and hard emission completely separable, with negligible contribution of the soft component above 3 keV, and vice versa. This justifies our choice of ignoring the data at E<3 keV in the data analysis.

**2. Spectral variability.** In the main text, we interpret the hardness ratio variability seen in the 3 – 10 keV band as due to a variable absorber. This is confirmed by time-resolved spectral analysis, but it is also easy to see visually by comparing the spectra from the four time intervals. Figure SI 4 shows the four spectra, normalized to have the same 7-10 keV flux. The similar spectral shape above 7 keV, and the strong deviations at lower energies are clear signatures of a variable absorber with column density in the range 1-3×$10^{23}$ cm$^{-2}$. The relativistic reflection component on the other hand can vary in flux, but not in spectral shape. The variability associated with the absorbing component helps in disentangling the absorption and reflection components, which are degenerate in the total, time-averaged 3 – 10 keV spectrum.

**3. Ionized absorber.** The absorption features present in the XMM-Newton and NuSTAR spectra between 6.7 and 8.3 keV are due to Fe XXV and Fe XXVI Kα and Kβ resonance lines. This implies the presence of a highly ionized absorber, with column density of the order of $10^{23}$ cm$^{-2}$ and with most of the metals completely ionised. As a consequence, its effect on the observed continuum is small. We consistently modeled this gas absorption component using the XSTAR code[29], with $N_H$~6×$10^{22}$ cm$^{-2}$, ionization parameter log$\zeta$=3.5±0.2, and iron abundance fixed to the best fit value for the reflection components. This choice may be not fully justified, given the different nature of the absorbing/reflecting gas. However, since the only significant effect of this component is through the iron absorption lines, the actual iron abundance is completely degenerate with the absorbing column density. Therefore, a change in iron abundance would be compensated by a change in $N_H$, leaving the global fit unchanged.

**4. Distant reflector.** The presence of a narrow iron line indicates the presence of a non-relativistic reflection component. Physically this could be associated with a distant torus or molecular clouds light-years away from the nucleus, and/or it could result from reflection from dense broad line region clouds closer in, at distances of a few light days. The variable Compton thin absorbers will also produce some line-emission in transmission. In our model fits, we assumed reflection consistent with distant Compton-thick material, however we demonstrate that the details of this non-relativistic reflection component are not important to our conclusions. Using REFLIONX[30] (appropriate for modeling reflection from dense Compton thick material) the best fit model finds the reflection component to be at an average level for Seyfert galaxies, with a 2-10 keV flux corresponding to that expected from a uniform Compton-thick screen covering ~1/3 of the solid angle as seen from the source. This reproduces the line and additional reflection continuum well. We also separated the narrow iron line and reflection continuum components, and obtained a similar fit, with all the relevant parameters unchanged within the errors. This indicates that our fits are insensitive to the details of the origin of this component.

A brief investigation into the origin of the distant reflector can be undertaken by replacing the reflection components in our fit with a model fixed to reproduce the emission observed in 2008 by XMM-Newton, which caught the source in a reflection-dominated state[31]. If the reflecting gas is located light-years away from the X-ray source, as in the standard ``torus'' of AGN Unified Models, this fit should reproduce the 'distant' reflector found our 2012 observations. We found however that this cannot reproduce our data, as it results in a reflection contiinuum too high at low energies (3-5 keV). This implies that the reflection continuum is not constant over years. This is not surprising in the case of NGC 1365: the obscuring gas along the line of sight is located at about the distance of the Broad Line Region, i.e. a few light days from the X-ray source, and it is likely that the same gas responsible for absorption is also the main contributor to reflection, as discussed in previous papers[15,31]. In this AGN there is no reason to expect reflection constant on time scales of years.

**5. Relativistic reflection model.** The best-fit relativistic reflection model is shown in Figure SI 5, together with the $\chi^2$ residuals for each instrument and each interval. Note the strong contribution of the relativistic reflection component, significantly shaping the total model in the 5-7 keV range, and becoming dominant in the 20-50 keV range. The constraints obtained for the disk/black hole parameters are: disk inclination $63^{+3}_{-20}$ degrees; black hole spin parameter $a^* \geq 0.84$. The best-fit iron abundance obtained for the disk is similar to previous work, Fe/solar~3, but is only weakly constrained by the data considered here. Both the covering fraction of the neutral absorber and the ionization state of the accretion disc were initially allowed to vary between the four intervals, but did not show any significant evolution within their statistical uncertainties. We therefore allowed these parameters to vary overall, but linked them between the four intervals. The covering factor obtained for the variable absorber is CF=$97.6^{+0.5}_{-0.4}$%, and the column densities obtained for the four time intervals are $N_H(1)=25.4^{+0.8}_{-1.4}\times10^{22}$ cm$^{-2}$, $N_H(2)=22.1^{+0.8}_{-1.2}\times10^{22}$ cm$^{-2}$, $N_H(3)=26.6^{+1.6}_{-1.4}\times10^{22}$ cm$^{-2}$, $N_H(4)=23.3^{+1.3}_{-1.1}\times10^{22}$ cm$^{-2}$. The continuum power law has a photon index $\Gamma=1.92^{+0.04}_{-0.26}$, and the global $\chi^2$ for the relativistic reflection model is 2117 for 2116 degrees of freedom. We also investigated whether the observed spectral variability could be explained by changes in photon index, but when allowed to vary between the time intervals the photon indices obtained were all consistent.

The results obtained here are consistent with those from Suzaku observations of NGC 1365 performed in 2008 and 2010 (REF 10). In particular, the first Suzaku observation of 2010 caught the source in a very similar absorption and flux state as in our new XMM-Newton + NuSTAR observation. However, the limited statistics, especially at energies higher than 10 keV made a comparison between the relativistic reflection and pure absorption scenarios impossible. As a consequence, in that work the relativistic model is assumed to be correct, and the possible systematic effects are not considered, analogously to several other spin measurements in bright AGN[2].

**6. Multiple absorbers with no relativistic reflection.** The final best fit absorption dominated model (with no relativistic disk reflection) requires three different absorbers, with variable covering factor and column density. Two components have column densities in the range $5\times10^{22}$-$3\times10^{23}$ cm$^{-2}$, with line of sight covering fraction $C_F$ of 50±10% and ~95%, respectively. The third component has $N_H$~4.5-$6.5\times10^{24}$ cm$^{-2}$. A formal fit implies the covering fraction of the Compton-thick absorber is $C_F$~0.45±0.05. We note that this value cannot be directly associated to a physical covering factor, as for the first two absorbers, because the model does not include the effect of Compton scattering. When this is taken into account (see Discussion Section) the estimate of intrinsic flux of the absorbed component is several times higher, and, consequently, the real CF is also higher.. The global $\chi^2$ for this model is 2131 for 2104 degrees of freedom. For this model, and for the variable absorber component in the relativistic reflection model, we assumed solar abundances. We checked that allowing the iron abundance to vary does not significantly change the best fit values and uncertainties of the relevant parameters.

**7. Statistical comparison of the two models.** The best fit $\chi^2$ for the relativistic model (2117/2116 degrees of freedom) suggests that this model is significantly preferred over the multiple absorber model, which has a $\chi^2$ higher by $\Delta\chi^2=14$ with 12 more free parameters. However, a quantitative statistical comparison between the two models would require complex Monte-Carlo simulations. We can, however, obtain a useful lower limit to the statistical preference for relativistic reflection by defining a ``parent'' model including all the components of both models (i.e. three variable absorbers, and relativistic reflection), and calculating the significance of each component through an F-test. The result is that while the extra absorbers are not required by the fit, the relativistic reflection is highly significant, with a rejection probability of $8\times10^{-5}$.

**Discussion.**
Here we discuss in more detail the physical implications of the three-zone multiple absorber model with no relativistic reflection. In particular, we concentrate on the implications of the Compton-thick absorption component, and we demonstrate that for a range of geometries bracketing the possibilities, straightforward physical arguments rule out the existence of this component.

The best-fit column density for the Compton-thick component in the four intervals correspond to an optical depth $\tau \sim 3$-$4$. We consider two extreme cases: (1) one single Compton-thick cloud which significantly attenuates the direct flux but results in minimal reprocessed radiation, and (2) a homogeneous torus covering >half of the solid angle, as seen from the source.

(1) If the absorption is due to a single, isolated cloud along the line of sight, the direct emission from the central source is depleted by a factor $\exp(-\tau_C) \sim 25$-$70$, while essentially no emission is scattered back into the line of sight. The corresponding intrinsic X-ray luminosity would then be very large(considering the covering factor for this component): $1.5$-$4 \times 10^{44}$ erg s$^{-1}$, to be compared with an estimated Eddington luminosity $L_{EDD} \sim 3 \times 10^{44}$ erg s$^{-1}$. The ratio between the flux in X-rays and in the [O III] $\lambda$ 5007 Å line, considered to be a rough estimator of the bolometric luminosity, would be $\sim 700$-$2000$. Average values for unobscured AGN are typically $\sim 10$ (REF 20) - much more than an order of magnitude lower. Such high values of Eddington and X-ray to [O III] ratios have never observed in other AGN, and are physically implausible.

(2) If the absorption is due to a toroidal, homogeneous absorber/reflector around the central X-ray source, multiple Compton interactions can scatter part of the radiation back to the line of sight, reducing the required luminosity of the central source. We used the PLCABS model[32] in XSPEC 12.3 to simulate this geometry, and we find that the expected primary radiation is "only" eight times higher than that implied by the relativistic reflection model, corresponding to a total 0.1-100 keV luminosity of $3.7 \times 10^{43}$ erg s$^{-1}$. This extreme scenario is obviously quite unlikely: a globally homogeneous absorber covering over half of the solid angle (as seen from the source, not to be confused with the line of sight covering factor, $C_F$) is hardly consistent with the observed $C_F \sim 40$-$50\%$ along the line of sight. Moreover, such an absorber should produce two observable effects: strong reprocessing in the infrared of the UV to hard X-ray primary emission, which conflicts with integral field IR spectroscopic observations[21]; and a strong narrow iron emission line.

The observed narrow iron line flux is inconsistent with the toroidal reflector, however demonstrating this quatitatively takes some more detailed analysis. In order to estimate the line equivalent width (EW), we used the *MyTorus* model[10], which self-consistently computes the continuum and line observed spectrum as seen through a gas torus covering half of the solid angle. Considering the *observed* fluxes of the direct and absorbed components, and correcting them for the effect of a $3$-$4 \times 10^{24}$ cm$^{-2}$ column density, we obtain a line of sight covering factor $C_F > 0.90$, and an observed EW of the narrow iron line of ~600-800 eV. This value exceeds the observed width (EW$_{obs}$~60) an order of magnitude.

The *MyTorus* code used for this estimate was specifically developed to predict the iron line strength from various absorbing/scattering geometries[22]. Applied to other AGN, similar calculations show that in some cases it is possible to produce a relatively weak iron line (EW~100 eV) even with the presence of a Compton-thick circumnuclear absorber/scatterer, once self-absorption of iron photons is taken into account. This may be the case, for example, for the AGN in MCG-6-30-15 (REF 33). However, applying the same calculation to NGC 1365 we still predict an excess of iron line by a factor of at least 10 with respect to the observed flux, unless we assume an extremely fine-tuned geometry consisting of an almost exactly edge-on, fully covering torus. This scenario is however not applicable to our case: the incomplete covering of the primary source implies that the absorber is *not* completely covering all lines of sight, thus leaving free directions for scattered iron photons. This structure would produce a total iron line flux more similar to that emitted by tori with intermediate inclinations, which in turn leads to the line over-prediction mentioned above. We note that the results would not change substantially if a spherical, instead of toroidal, geometry is assumed.

Another possible scenario, is that the Compton-thick absorber is extremely compact and moderately ionized (log $\xi$ ~2). Such a structure would produce neither strong infrared reprocessing, nor strong iron line emission However, in order to have the appropriate ionization state, the absorber needs to be located within a few $10^{15}$ cm from the central source, and have a density higher than $10^9$ cm$^{-3}$. Such a configuration is probably unphysical, however it can also be ruled out based on observational evidence: the ionized gas would produce a soft X-ray luminosity L>$10^{42}$ ers s$^{-1}$, to be compared with the 0.3-2 keV unresolved X-ray luminosity L<$10^{40}$ erg s$^{-1}$, as measured by the Chandra observatory[18]. We checked that a strong excess with respect to the observed Chandra emission would be detected even in the case the warm gas is interior to the Compton-thin absorbers.

Finally, we note that both scenarios discussed here imply a large (>90%) line of sight covering factor $C_F$ of the Compton-thick absorber. In all cases the absorber must be located at least at hundreds (in the case of highly ionized gas) or thousands of $R_G$. Since the X-ray source has a linear dimension of only a few $R_G$, such a configuration requires a very sharp-edged, stable set of clouds in an extremely fine-tuned spatial distribution. This further adds to the previous arguments against this model.

Based on all the observational constraints and the physical arguments discussed above, we conclude that the multiple absorber model cannot provide a physically acceptable explanation for the hard X-ray excess observed by NuSTAR.

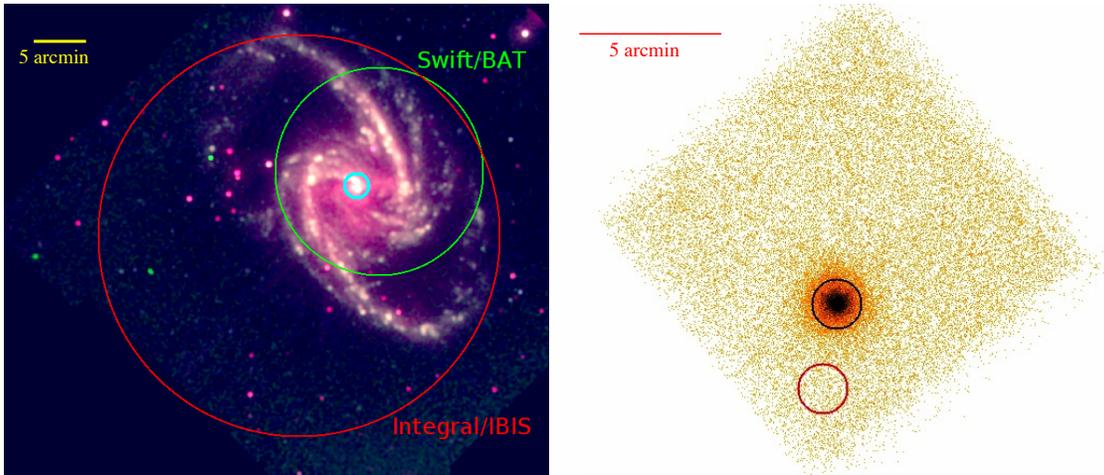

**Figure SI 1.** Left panel: UV image of NGC 1365, obtained with the XMM-Newton Optical Monitor. The circular regions show the spatial resolution (90% encircled area) of NuSTAR (the small, light blue central circle), Swift/BAT and Integral/IBIS. Right panel: image from the NuSTAR FPMA module, with the circular regions used for source and background extraction.

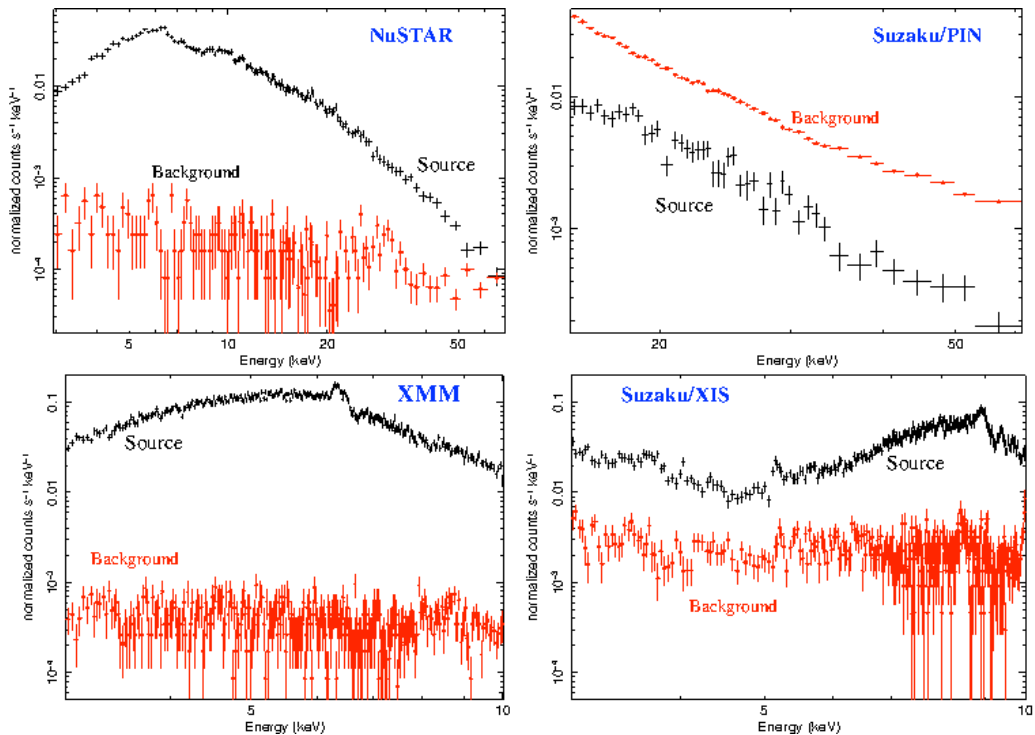

**Figure SI 2. Left panels:** source + background spectra from the NuSTAR FPMA module for the 78 ks (130 ks elapsed time) observation of NGC 1365 in July 2012, and from the simultaneous XMM-Newton observation,. Right panels: same for a Suzaku observation of NGC 1365 (instruments PIN in the 15-70 keV range, and XIS

in the 3-10 keV range), performed in 2010 (REF 10). The two observations have the same duration and the average fluxes, within a few percent.

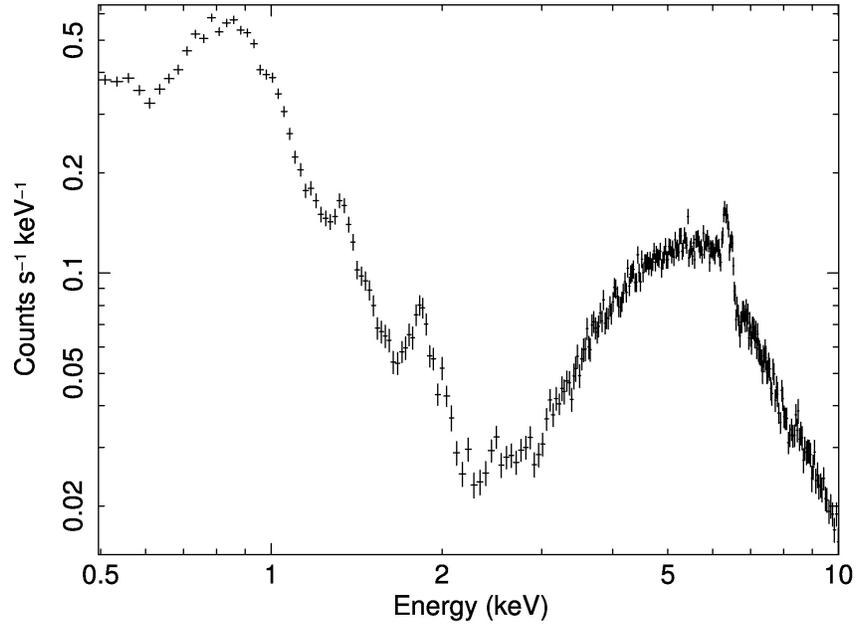

**Figure SI 3.** 0.5-10 keV total XMM-Newton spectrum of NGC 1365, showing the spectral separation between the thermal component below 3 keV, due to extended gas, and the hard AGN emission above 3 keV.

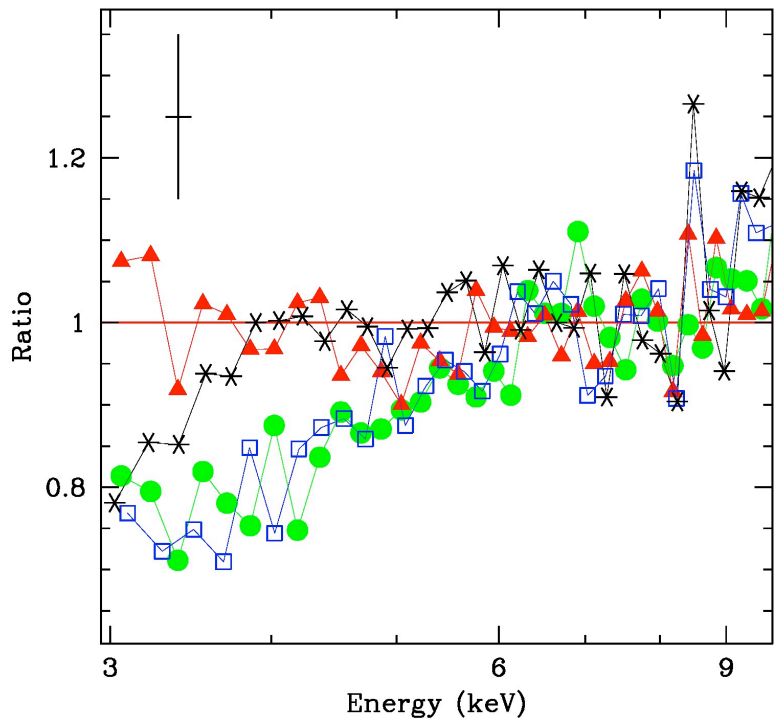

**Figure SI 4** XMM-Newton 3-10 keV spectra of the four time intervals discussed in the main paper. The spectra have been normalized in order to have the same flux in the 7-10 keV range. A typical errorbar is shown in the upper left corner.

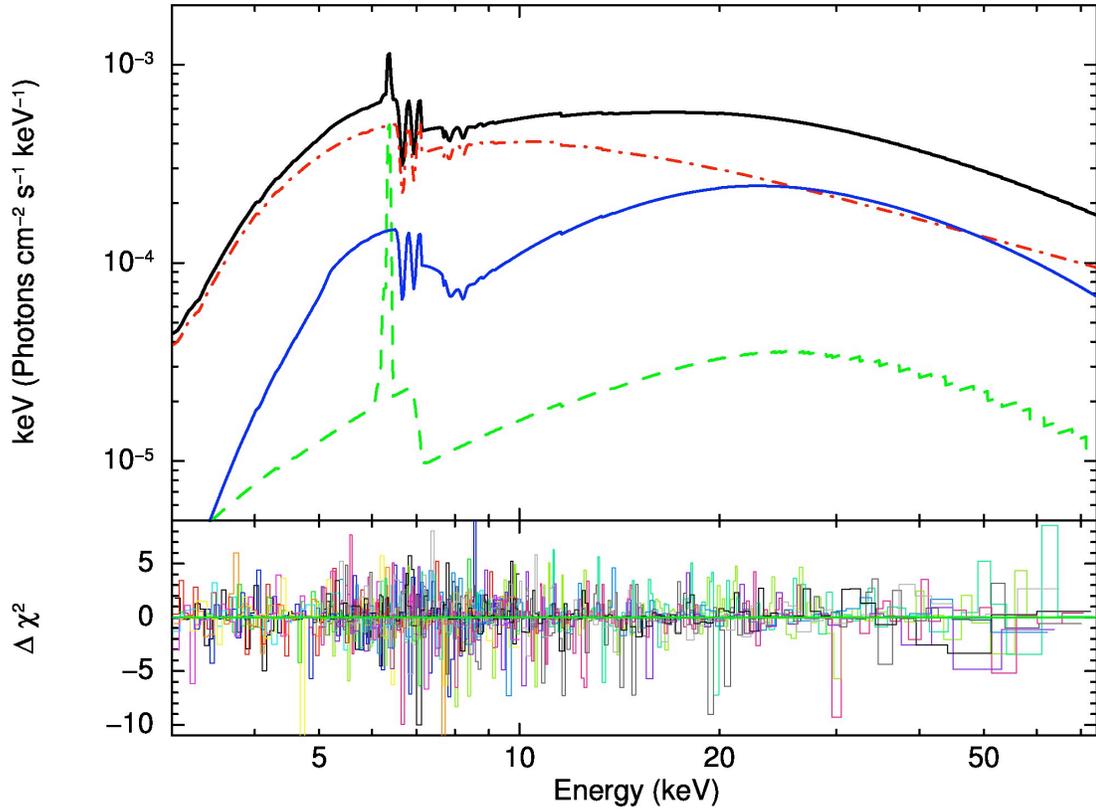

**Figure SI 5** Upper panel: Best fit model from the joint XMM-Newton + NuSTAR spectral analysis. The total model and the main components are shown: the absorbed primary power law (red, dot-dashed line), the relativistic reflection (blue, continuous line), the reflection from a distant screen (green, dashed line). Lower panel: $\chi^2$ residuals for all the XMM-Newton and NuSTAR spectra from the four time intervals.